\documentclass[aps,prl,amsmath,amssymb,superscriptaddress,notitlepage]{revtex4-1} 

\usepackage{graphicx}
\usepackage{amssymb}
\usepackage{amsmath}
\usepackage{dsfont}
\usepackage{bm}
\usepackage{mathrsfs}
\usepackage{times}
\usepackage[utf8]{inputenc} 
\usepackage{gensymb} 
\usepackage{xcolor}

\newcommand{\be}{\begin{equation}}
\newcommand{\ee}{\end{equation}}

\begin{document}
\title{Visco-elastic Cosmology for a Sparkling Universe?}

%\author{S.\ Mancas}
%\affiliation{Mathematics Department, Daytona College of Arts \& Sciences, Daytona Beach Campus, USA}
\author{G.\ Rousseaux}
\email{Corresponding author: germain.rousseaux@univ-poitiers.fr}
\affiliation{Institut Pprime, UPR 3346, CNRS-Universit\'{e} de Poitiers-ISAE ENSMA, TSA 51124, 86073 Poitiers Cedex 9, France}

\author{S. C. \ Mancas}
\email{mancass@erau.edu}
\affiliation{Department of Mathematics, Embry-Riddle Aeronautical University,\\ Daytona Beach, FL. 32114-3900, U.S.A.}

\begin{abstract}
We show the analogy between a generalization of the Rayleigh-Plesset equation of bubble dynamics including surface tension, elasticity and viscosity effects with a reformulation of the Friedmann-Lema\^itre set of equations describing the expansion of space in cosmology assuming a homogeneous and isotropic universe. By comparing both fluid and cosmic equations, we propose a bold generalization of the newly-derived cosmic equation mapping three continuum mechanics contributions. Conversely, the addition of a cosmological constant-like term in the fluid equation would lead also to a new phenomenology.
\end{abstract}
  
  \maketitle

\section*{}
Banerjee et al.  proposed recently a new model of cosmic dynamics to solve the enigma of the so-called dark energy in order to cope with the observed expansion of the Universe with an ever-accelerating rate \cite{Banerjee}. They modified the first Friedmann-Lema\^itre equation with a quartic correction in the fourth inverse power of the scale factor \cite{Friedmann, Lemaitre}. Their procedure is compatible with the picture of a universe riding on an expanding bubble which is being inflated by the dark energy in an extra dimension. In addition to our 3+1 Universe placed on the surface of the bubble like a membrane, there must be another ``parent" 5D Universe. According to Banerjee et al. \cite{Banerjee}, {\it ``the cosmology we see as 4D observers is not due to vacuum energy, but rather arises as an effective description on a dynamical object embedded in a higher dimensional space...all processes on the shellworld will be like shadows of processes taking place in 5D involving much larger energies".} Can the Universe be described literally by a bubble dynamical equation in a kind of cosmic fluid? If so, would this ``sparkling Universe" feel viscous damping of any kind? The energy-momentum tensor $T_{\mu \nu}$ in General Relativity does not feature dissipative effects usually so a generalization of the Friedmann-Lema\^itre equations is not obvious. Here, we show, thanks to an analogy with the extended Rayleigh-Plesset equation which describes the radial dynamics of a bubble in a classical viscous fluid with surface tension \cite{Rayleigh, Plesset, Minnaert, NN, Poritsky, FM, Chaline}, what would be the corrective term. A formal equivalence between the motion of an inviscid fluid in a capillary tube and the Friedmann-Robertson-Walker cosmological equations was discussed a few years ago in \cite{BS}. Another interesting analogy with a Coulomb system has recently been reported by Kolomeisky and bears some ressemblances with our proposal \cite{Kolomeisky} as well as a discussion by one of us who established a partial connection between the Rayleigh-Plesset and Friedmann-Lema\^itre systems \cite{Mancas1, Mancas2, Mancas3}. Other partial analogies were introduced very recently with some similitudes with either equilibrium beach profiles or freezing lakes by Faraoni \cite{Faraoni1, Faraoni2}.\\

Both Friedmann-Lema\^itre equations may be written in the following form (including the cosmological constant term \cite{EPJH}) amenable to the study of cosmic phenomena on a time-varying spatial domain \cite{Friedmann, Lemaitre, KR}
\begin{equation} 
\left(\frac{\dot{a}}{a}\right)^2+\kappa c^2\left(\frac{1}{a}\right)^2=\frac{8\pi G}{3}\rho+\frac{\Lambda c^2}{3},
\label{FL1}
\end{equation} 
\begin{equation} 
\frac{\ddot{a}}{a}=-\frac{4\pi G}{3}\left(\rho+\frac{3p}{c^2}\right) +\frac{\Lambda c^2}{3}.
\label{FL2}
\end{equation} 

The first equation is derived from the $00$ component of Einstein's field equations  assuming an isotropic and homogeneous universe, while the second one, known as the acceleration equation, was derived from the first one  together with the trace of the same equations. The dynamical variable of interest is the scale factor $a (t)$ which is a function of the co-moving time $t$. One also defines $H=\dot{a}/a$ (where the dot denotes the time derivative) the Hubble parameter that is accessible with astronomical observations. $G$ is Newton's gravitational constant, $\Lambda$ is the Einstein's infamous cosmological unknown constant which is related to the vacuum pressure, and may take any sign, and $c$ is the speed of light in vacuum. The term $\kappa/a^2$ is the spatial curvature in any time-slice of the universe with $\kappa$, the spatial curvature index, which  takes the values of $0$ or $\pm 1$ for  flat or closed/open universe respectively.  The pressure  $p$ and energy density  $\rho$ are  functions of time and are related by a constitutive thermodynamics law,  which is the equation of state for a barotropic fluid
\begin{equation}
p=(\gamma-1) \rho c^2.
\end{equation} 
$\gamma$ is the adiabatic index which for the cosmological matter takes the values of $0$ for vacuum, $1/3$ for domain walls, $2/3$ for cosmic strings, $1$ for dust, $4/3$ for radiation, $5/3$ for mono-atomic perfect gas, and $2$ for Zeldovich stiff matter \cite{Zeldo}. The non positive values of $\gamma$ correspond to dark energy, and have been studied in context of phantom matter \cite{Dab}.

By eliminating  the density  $\rho$ among Friedmann-Lema\^itre equations, the acceleration equation  takes the equivalent form
\begin{equation} 
2\left(\frac{\ddot{a}}{a}\right)+ \left(\frac{\dot a}{a}\right)^2+\kappa c^2\left(\frac{1}{a}\right)^2 =-\frac{8 \pi G}{c^2}p+\Lambda c^2.
\label{FL3}
\end{equation}

In general, equations (\ref{FL1}) and  (\ref{FL3}) are preferred over (\ref{FL1}) and  (\ref{FL2}), because if we know the scale factor $a(t)$, the pressure  $p$ and energy density  $\rho$ are  found directly, and this is what exactly  Lema\^itre  \cite{Lemaitre} has used in his work apart from the $\kappa c^2$ factor which he set it as $1$.

By letting $w=\gamma-1$, then we have $w<-1$ for  dark energy, $w=-1$ for vacuum, $w=-2/3$ for domain walls, $w=-1/3$ for cosmic strings, $w=0$ for dust, $w=1/3$ for radiation, $w=2/3$ for a mono-atomic perfect gas, and  $w=1$ for stiff matter.  Eliminating the pressure via the  constitutive thermodynamics law  $p=w \rho c^2$  between  (\ref {FL1}) and (\ref{FL3}) yields an equivalent form of the acceleration equation
\begin{equation}
\frac{\ddot{a}}{a}+ \frac 32 \left(\frac{\dot a}{a}\right)^2+\frac 32 \kappa c^2\left(\frac{1}{a}\right)^2 =\frac{4 \pi G}{3}(2-3 w)\rho+\frac 5 6\Lambda c^2.
\label{FL2bis}
\end{equation}

The conservation of energy equation taking into account the same law
\begin{equation}
\dot \rho+3\frac{\dot a}{a}\left(\rho+\frac{p}{c^2}\right)=0
\label{cons}
\end{equation} integrates to 
\begin{equation}\label{den}
\rho=\frac{3}{8 \pi G}\frac{c_w}{a^{3(w+1)}},
\end{equation} where $c_w$ is an arbitrary  constant, and $\rho$ is the total energy density  which may include in general  various  terms such as  vacuum density $\rho^{vac}$, radiation density $\rho^{rad}$, or density of dust $\rho^{dust}$. Assuming a spherical volume $V= \pi^2 a^3$ of constant mass $M$ the conservation of energy  (\ref{cons}) has an interesting representation
\begin{equation}\label{cons2}
\frac{d}{dt}\left(\rho V\right)+\frac{p}{c^2}\frac{dV}{dt}=0,
\end{equation} which shows that the variation of total energy  plus the work done by the radiation is zero.

By substituting (\ref{den}) into (\ref{FL2bis}), we obtain the general acceleration equation where the scale factor  depends only on the curvature and cosmological constant
\begin{equation}
\frac{\ddot{a}}{a}+ \frac 32 \left(\frac{\dot a}{a}\right)^2+\frac 32 \kappa c^2\left(\frac{1}{a}\right)^2 =\left(1-\frac 3 2 w\right)\frac{c_w}{a^{3(w+1)}}+\frac 5 6\Lambda c^2.
\label{equis}
\end{equation}

Now let us identify some density terms in this equation. For $w=-1/3$, or $\gamma=2/3$ we have the cosmic strings density $\rho^{str}=\frac{3}{8 \pi G} \frac{c_{-1/3}}{a^2}$, so the curvature term is equivalent   to this  density.  Cosmic strings are hypothetical 1D topological defects which may have formed during a symmetry breaking phase transition in the early universe, were first introduced by Kibble in the late 70's, and were the subject of search for gravitational waves  from cosmic string cusps in data collected by the LIGO and Virgo gravitational wave detectors between 2005 and 2010  \cite{Kibble}. Notice that $c_{-1/3}$ corresponds to $\kappa  c^2$, while  the strings pressure  is given by $p^{str}=-\frac{\kappa}{8 \pi G}\frac{c^2}{a^2}$. For closed universe, the pressure is negative while open universe has positive pressure.

The vacuum density term is obtained for  $w=-1$, or $\gamma=0$ and  for which $\rho^{vac}=\frac{3}{8 \pi G}c_{-1}=\frac{\Lambda c^2}{8 \pi G}$,  and implies that the integration constant is related to the cosmological constant by $c_{-1}=\frac{\Lambda c^2}{3}$. Thus, the cosmological constant is equivalent to a non-zero vacuum energy. Notice that  $c_{-1}$ appears exactly in the last two terms of (\ref {FL1}) and (\ref {FL2}), and it might be the only reason of why  the universe is expanding, as de Sitter was depicted in his famous  caricature from $1930$ when he was happily blowing air in an expanding bubble: {\it ``But who inflates the ball? What causes the Universe to expand, or swell up? That's what lambda does -- a different answer cannot be given"} \cite{Icke}. For this case, the vacuum pressure is given by $p^{vac}=-\frac{\Lambda}{8 \pi G}c^4$, where 
$T^{vac}_{\mu\nu}=p^{vac}~g_{\mu\nu}.$ Thus, for $\Lambda>0$ we have a  negative pressure which agrees with what de Sitter was proposing. 

Other terms that can be included in the analysis are the radiation density  $\rho^{rad}=\frac{3}{8 \pi G}\frac{c_{1/3}}{a^4}$ which corresponds to $w=\frac 1 3$ for which $\gamma=4/3$ with radiation pressure given by $p^{rad}=\frac{c_{1/3}}{8 \pi G}\frac{c^2}{a^4}$. For a universe containing  dust that corresponds to $w=0$ or $\gamma=1$, we have  a dust density  $\rho^{dust}=\frac{3}{8 \pi G}\frac{c_0}{a^3}$ and zero pressure $p^{dust}=0$.

Can the Universe be described literally by a bubble dynamical equation in a kind of cosmic fluid? The radial dynamics for an incompressible flow of a bubble in a classical fluid is described by the Extended Rayleigh-Plesset (ERP) equation due to the subsequent works of Rayleigh, Plesset, Minnaert, Noltingk \& Neppiras and Poritsky \cite{Rayleigh, Plesset, Minnaert, NN, Poritsky, FM, Chaline}
\begin{equation}
\frac{\ddot{R}}{R} + \frac{3}{2} \left(\frac{\dot{R}}{R}\right)^2+ \frac{-\Delta P(t)}{\rho_L}\left(\frac{1}{R}\right)^2 + \frac{2\sigma}{\rho_L} \left(\frac{1}{R}\right)^3 + 4\nu \left(\frac{\dot{R}}{R^3}\right) =0,
\label{ERP}
\end{equation}
where R(t) is the time-evolving radius of the bubble, $\rho_L$ the liquid density, $\nu$ the liquid kinematic viscosity, $\sigma$ the surface tension of the bubble-liquid interface.
$\Delta P(t)= P_{in}(t) - P_{out}(t)$ is the pressure drop between the uniform pressure in the bubble and the external pressure in the liquid at infinity (hydrostatic and sound field for example \cite{FM}). In the stationary limit, the ERP equation leads to the Young-Laplace equilibrium equation $\Delta P=2\sigma /R=\sigma C$ where $C$ is the bubble mean curvature. Equations (\ref{equis}) and (\ref{ERP}) are identical for a  cosmic  fluid containing  dust  with zero cosmological constant and curvature given by $\kappa=-\frac 2 3 \frac{\Delta P}{c^2 \rho_L}$, and for an inviscid water bubble  with surface tension given by $\sigma=-\frac 1 2 c_0 \rho_L$.

Can we rewrite the Friedmann-Lema\^itre system of equations in a form analogous to the ERP equation? The answer is positive. Indeed, we can express the energy density as a function of the scale factor and the cosmological constant term thanks to the  equation
\begin{equation} 
\rho=\left(\frac{\Lambda c^2}{3}-\frac{\ddot{a}}{a}\right)\frac{3}{4\pi G(1+3w)},
\label{rho}
\end{equation} 
and by using this expression in Friedman equation we obtain 
\begin{equation} 
\frac{\ddot{a}}{a}+ \left(\frac{1+3w}{2}\right) \left(\frac{\dot{a}}{a}\right)^2 + \left(\frac{1+3w}{2}\right)\kappa c^2\left(\frac{1}{a^2}\right) \\ -\left(\frac{1+w}{2}\right)\Lambda c^2=0.
\label{CFL}
\end{equation}
For  mono-atomic perfect gas $w=2/3$  which corresponds to $\gamma=5/3$  the combined Friedmann-Lema\^itre (CFL) equation (\ref{CFL}) becomes
\begin{equation} 
\frac{\ddot{a}}{a} + \frac{3}{2} \left(\frac{\dot{a}}{a}\right)^2 + \frac{3}{2}\kappa c^2\left(\frac{1}{a}\right)^2- \frac{5}{6}\Lambda c^2=0.
\label{GCFL}
\end{equation}

In absence of both surface tension and viscosity, the Rayleigh-Plesset equation is strictly identical to the CFL equation for a mono-atomic gas where the enthalpy drop $\Delta P/\rho_L$  is analogous to $-3/2\kappa c^2$ with zero cosmological constant. This  explains that positive pressure difference accounts for open universe when inner pressure is larger than outer pressure, while negative difference accounts  for closed universe.

By comparing the ERP equation with the CFL equation, we propose the following  generalization of the CFL equation
\begin{equation} 
\frac{\ddot{a}}{a}+ \chi \left(\frac{\dot{a}}{a}\right)^2 + \chi~ \kappa c^2\left(\frac{1}{a^2}\right) -\left(\frac{1+\chi}{3}\right)\Lambda c^2+ \alpha \left(\frac{1}{a}\right)^3+\beta \left(\frac{1}{a}\right)^4+\delta\left(\frac{\dot{a}}{a^3}\right)=0,
\label{GCFL+}
\end{equation}  
where $\chi=\frac{1+3w}{2}.$ This takes into account the quartic correction term $1/a^4$ of Banerjee which was calculated due to the back-reaction of bulk matter \cite{Banerjee} together with the term $\dot a/a^3$  which corresponds to the  viscous term in the RP equation. To show the analogy, we need to explain  that  the total energy density $\rho$  contains  both  dust  $ \rho^{dust}$  and radiation $\rho^{rad} $ terms which  correspond to the inverse cubic and quartic terms. 
From (\ref{cons2}), and for  a spherical volume space of volume $V= \pi^2 a^3$ and constant mass $M=4\pi^2 a^3 \epsilon$, but non constant density $\epsilon$  we have
\begin{equation}\label{red}
d(\rho a^3)+\frac{3p}{c^2}a^2 da=0.
\end{equation}
Taking into the account the relation among dust and radiation $\rho=\epsilon+\frac{3p}{c^2}$, where $\rho$ is the total density, $3p/c^2$ is the density of radiation and the rest $\epsilon$ is the density of dust, then (\ref{red}) reduces to
\begin{equation}
d(pa^3)+pa^2 da=0.
\end{equation}
Solving this equation for the pressure as a function of the scaling factor, we obtain the pressure  $p=\frac{A}{a^4}$, where $A$ is some constant.  Consequently, we obtain  the total density $\rho=\frac{M}{4\pi^2 a^3}+\frac{3A}{c^2 a^4}$ which can be written as $\rho=\frac{\alpha}{a^3}+\frac{\beta}{a^4}.$ This analogy is consistent  with our values of $w$, for dust $w=0$ so $p^{dust}=0$, thus $\rho=\rho^{dust}=\epsilon$, while for radiation $\epsilon=0$ which implies $p^{rad}=\rho^{rad} c^2/3$, i.e., $w=1/3$. 
Conversely, a generalization of the ERP equation could take into account both a different constitutive relation \cite{FM, Chaline} and a cosmological-like term. The latter is equivalent to the square of an oscillation frequency since one gets $\ddot{R}+\iota^2R=...$ if the cosmological constant is positive. Otherwise, one gets a damping equation $\ddot{R}-\hat \iota ^2R=...$ with a damping time scale $\hat \iota$ for a negative sign of the cosmological constant. Of course, one could drop for the sake of clarity the cosmological constant term on the cosmology side, which causes a problem in the analogy, but we feel that maybe fluid mechanics could lead us to an interpretation of the cosmological constant by probing some physical effects from continuum mechanics e.g. cavitation with its negative pressure \cite{FM}, chemical reactions, dissolution or thermal heating which would maybe explain its true nature. We left this question to the readers since we do not want to hide some interesting questions raised by the analogy. Similarly, the inclusion of a viscous term on the cosmology side is not necessary, but the true interest of the analogy is precisely to point to the good mathematical expression with its unknown coefficient that more accurate models or numerical simulations and observations could one day determine exactly. As pointed out in the introduction, Einstein's equations are at loss when dealing with dissipative terms, so maybe our heuristic model can be of some help to future explorations. As a final remark worthy of further developments, the cosmological constant term $\Lambda$ (when positive) is a source term for the gravitational field inducing repulsion: in the Newtonian limit, the evolution of the self-gravitating continuum is given by $\nabla \cdot {\bf g}=\Lambda - 4\pi G \rho$ for the gravitational field strength ${\bf g}({\bf x}, t)$. Hence, in fluid mechanics, maybe one could relax the conservation of mass $\nabla \cdot {\bf u}=0$ for the velocity field ${\bf u}({\bf x}, t)$ by adding a mass source term that would contribute to the Rayleigh-Plesset equation with an additional term.

By balancing each term of the generalized cosmic equation \`a la Rayleigh-Plesset with each of the other terms using orders of magnitude, we recover immediately the well-known scaling laws for the scaling factor $a(t)\approx t$ for curvature, $a(t)\approx t^{2/3}$ for dust, and  $a(t)\approx t^{1/2}$ for radiation. We point out that the newly-added viscous term leads to a similar scaling compared to radiation namely $a(t)\approx t^{1/2}$ and a linear scaling  $a(t)\approx t$ when compared to dust.

Concerning the quartic correction of Banerjee et al. \cite{Banerjee}, the ``bubble with shell" model (where the gravitational back-reaction of bulk matter is identified as the source of an effective energy density with a radiation equation of state on the shell) is an exact analogue of the Elastic Extended Rayleigh-Plesset ($EERP$) equation:
\begin{equation}
\frac{\ddot{R}}{R} + \frac{3}{2} \left(\frac{\dot{R}}{R}\right)^2+ \frac{-\Delta P(t)}{\rho_L}\left(\frac{1}{R}\right)^2 + \\ \frac{2\sigma}{\rho_L} \left(\frac{1}{R}\right)^3 + \frac{-Yh^2}{\rho_L}\left(\frac{1}{R}\right)^4 + 4\nu \left(\frac{\dot{R}}{R^3}\right) =0
\label{eERP}
\end{equation}
with an additional bending pressure term $p_B=Yh^2/R^2$ of the thin elastic outer shell coating the bubble \cite{MKT}, where $Y$ is the elastic modulus of the shell and $h$ its thickness. By analogy of equations  (\ref{GCFL+}) and (\ref {eERP}) the surface tension $\sigma$ corresponds to the constant for  dust $c_0$, while the elastic term corresponds to the radiation constant $c_{1/3}$. The cubic term due to surface tension in equation (\ref{eERP}) should not be confused with the quartic term associated to elasticity as is well known in continuum mechanics, whereas the physical interpretation in astrophysics is the following: the fluid-like cubic term corresponds to the usual presence of dust whereas the solid-like quartic term has to be interpreted directly to the usual contribution of radiation or to the effect of 5D matter confined to the shell in the model of Banerjee et al. \cite{Banerjee} that {\it ``identify the gravitational backreaction of bulk matter as the source of an effective energy density with a radiation equation of state on the shell"}.\\

As a perspective, more complicated models of bubble encapsulation exist in the literature, and the interested reader will have a look to all the possible extensions to cosmology by taking inspiration of the works discussed in the review \cite{DB}. Dynamical systems theory allows to solve (using a phase-space analysis) the Rayleigh-Plesset equation and its variants, similar studies should be undertaken for the analogue cosmological model proposed in this work. The inclusion of a secondary dilatational/bulk viscosity term \`a la Weinberg $\dot{R}/R$ would be an interesting add-on since our viscous correction is only of the shear type $\dot{R}/R^3$ \cite{Weinberg}. In presence of a shell of a given thickness $h$, the dimensionless ratio of the thickness to the radius multiplies the usual viscous term leading to another shell dissipative term $h\dot{R}/R^4$ featuring a microscopic scale $h$ even in more complex models \cite{Marmottant05}. Numerical simulations and comparison with observational data could lead to the values of the $\alpha$, $\beta$ and $\delta$ coefficients. Our generalized visco-elastic-capillary Friedman-Lema\^itre model is straight forward, but the history of science tells us about many examples of the use of analogies (see the recent example of Analogue Gravity for instance \cite{Barcelo18, PRL2020}) to construct heuristic models which have been later on recognized as discovery tools. It is savoury to notice that Lema\^itre and other founders of cosmology pondered about the existence of so-called phoenix universes whose radius varies cycloidally with alternance of Big Bangs and Big Crushes \cite{HK, Zeldo}: a typical bubble-like behavior!\\

{\it Acknowledgements:} GR thanks Jennifer Chaline for providing him with the essential knowledge on the Rayleigh-Plesset equation based on her Ph.D. thesis work. He also acknowledges discussions with M. Baudoin and P. Marmottant.

{\it Declaration of Interests:} The authors report no conflict of interest.

{\it Data Sharing:} Data sharing is not applicable to this article as no new data were created or analyzed in this study.

\end{document}